\documentstyle[11pt,newpasp,twoside,epsfig]{article} 
\def\ApJ{{\it Astrophys. J.} } 
\def\ApJL{{\it Astrophys. J. Letters} } 
 
\def\ApP{{\it Astropart. Phys.} } 
\def\AA{{\it Astron. \& Astroph.} } 
 
\def\AAL{{\it Astron. \& Astroph. Letters} } 
\def\JGR{{\it Journ. of Geophys. Res.}} 
 
\def\PhFl{{\it Phys. of Fluids} } 
 
\def\PRD{{\it Phys. Rev.} {D} } 
\def\PRL{{\it Phys. Rev. Letters} } 
\def\Nature{{\it Nature} } 
\def\MNRAS{{\it Month. Not. Roy. Astr. Soc.} } 
 
\def\ARAA{{\it Annual Rev. of Astron. \& Astrophys.} }

\def\etal{{\it et al.~} } 

\def\simle{\lower 2pt \hbox {$\buildrel < \over {\scriptstyle \sim }$}} 
\def\simge{\lower 2pt \hbox {$\buildrel > \over {\scriptstyle \sim }$}} 

\begin{document} 

\title{Cosmic Rays in Clusters of Galaxies} 

\author{Peter L. Biermann} 
\affil{Max-Planck-Institut f{\"u}r Radioastronomie and University 
of Bonn, Bonn, Germany} 
\author{Torsten A. En{\ss}lin} 
\affil{Max-Planck-Institut f{\"u}r Astrophysik, M{\"u}nchen, Germany} 
\author{Hyesung Kang} 
\affil{Pusan National University, Pusan 609-735, Korea} 
\author{Hyesook Lee} 
\affil{Chungnam National University, Daejeon 305-764, Korea; and 
Max-Planck-Institut f{\"u}r Radioastronomie, Bonn, Germany} 
\author{Dongsu Ryu} 
\affil{Chungnam National University, Daejeon 305-764, Korea} 

% 
% version of Oct 17, 2002, last stage of revision (PLB) 
% 
% sequence of revision: PLB -> DR -> TE -> HL -> HK -> PLB. 
% emails: 
% kang@msi.umn.edu,ensslin@MPA-Garching.MPG.DE, 
% zoarlee@hanmail.net,ryu@msi.umn.edu, 
% plbiermann@mpifr-bonn.mpg.de,J.Cobb.Biermann@t-online.de 
% 
% comments by Phil Kronberg and editor Stu Bowyer 
% 
% Philipp Kronberg <kronberg@lanl.gov> 
% 
% at the end to Stu Bowyer and hwangcy@astro.ncu.edu.tw 
% stubowyer@netscape.net (stuart bowyer) 
% Stu Bowyer <bowyer@albert.ssl.berkeley.edu> 
% 

\abstract 
{We argue that clusters of galaxies have an 
intergalactic medium, which is permeated by strong magnetic fields and 
also has a contribution of pressure from cosmic rays. These two 
components of total pressure are probably highly time dependent, and 
range probably between 1/10 of the gas pressure up to equipartition 
between gas pressure and the sum of the two other components. Radio 
galaxies are likely to provide the main source for both magnetic fields 
and cosmic rays. In this concept it becomes easy to understand the 
occasional mismatch between the total mass inferred from the assumption 
of hydrostatic equilibrium derived purely from gas, and the total 
mass derived from lensing data. We also suggest that the structure 
and topology of the magnetic field may be highly inhomogeneous - at least 
over a certain range of scales, and may contain long twisted filaments of 
strong magnetic fields, as on the Sun. The analogy with the interstellar 
medium may be fruitful to explore further, where we do not know where 
magnetic fields come from, but suspect that the cosmic rays derive from 
supernova explosions. In such an analogy it becomes useful to refer to 
``radio galaxy explosions" in clusters of galaxies. A full scale 
exploration of all the implications, especially of the notion that 
occasionally complete equipartition may be reached, is a task for the 
future.}

\section{Introduction} 

For a long time it has been recognized that clusters of galaxies have 
intergalactic gas, which actually dominates in baryonic mass over the 
stars in galaxies; this gas is enriched in heavy elements. The 
assumption of hydrostatic equilibrium can be used to infer in the case 
of a nearly spherical cluster the total gravitating mass of a cluster; 
this inferred mass can in turn be tested with strong lensing 
observations which also give the gravitating mass; for spherical 
clusters with high accuracy. It had been believed for some time that 
this intergalactic gas in clusters appeared to be much simpler than 
interstellar gas in the sense that magnetic fields and cosmic ray 
particles did not seem to matter much for the overall energetics, the 
overall pressure, and the emission. We can refer to the ensemble of 
magnetic fields and energetic particles, i.e. cosmic rays, as the 
``nonthermal component". In recent years this view has begun to be 
demonstrated to be false, and we now have to take these two other 
components into account, leading to the same questions and uncertainty 
as for the interstellar medium, for which we are just beginning to 
appreciate the difficulties in understanding its physics, as, for 
instance, in searching for the origin of the interstellar magnetic 
field. 

Here we discuss what is known today about the nonthermal 
component of the intergalactic medium in clusters. We outline a range of 
possibilities, and tasks. These questions are all the more important 
since we now suspect that the missing baryonic mass in the universe is 
all in gas, usually outside of clusters of galaxies, in groups, filaments 
and sheets of the galaxy distribution. 

The order of the paper is as follows: 1) First we outline some basic 
parameters of the intergalactic medium in clusters, in this introductory 
section. 2) We discuss the recent detection of rather strong magnetic 
fields in clusters. 3) We discuss cosmic rays in clusters, as derived 
from normal galaxies and starburst galaxies. 4) Then we discuss the cosmic 
rays from accretion shocks. 5) Mergers between clusters and structural 
rearrangements of un-relaxed clusters can also produce shock waves and so 
cosmic rays. 6) Finally, cosmic rays from radio galaxies can surely 
provide sufficient energy to provide cosmic rays, and also magnetic fields 
to clusters of galaxies, and do so to the stability limit of the medium. 
7) Then, we describe the attempts to understand where cosmic magnetic 
fields come from. 8) We discuss explosions into a 
homogeneous medium, so as to describe the effective cooling in 
either the interstellar medium or intergalactic medium, subject to the 
buffeting of supernova explosions or radio galaxy explosions. We discuss 
to what limit this analogy may help to understand the physics of these 
two media. 9) Also, the recent XMM-Newton findings illustrate the 
necessity to consider the heating from radio galaxies in the cores of 
clusters. 10) Then, we describe some cosmological simulations that 
show what happens with magnetic fields. 11) We then explore the topology 
of magnetic fields in a medium subject to continuous excitement of 
turbulence, again as an example for both the interstellar and the 
intergalactic medium. 12) We furthermore discuss the two 
sources of leptons so as to predict the lepton energy distribution, 
critical to understand the inverse-Compton and synchrotron emission in 
clusters; leptons are expected both from the primary acceleration of 
electrons starting from the thermal tail, as well as from pion decay 
following p-p collisions between energetic protons and the medium itself. 
13) We then outline some predictions as regards the various possible 
limits of the model proposed here, namely that radio galaxies dominate 
both the source of magnetic field and the source of energetic particles, 
the cosmic rays. 14) We conclude with some final summary of the present 
status of this field, and define some tasks for the future.

\subsection{Basic properties of clusters} 

First we outline some basic parameters of the intergalactic medium in 
clusters, in this introductory section. 

Clusters of galaxies cover an enormous range of properties, some 
probably still unrecognized. But ``typical" values for the most massive 
clusters are by order of magnitude: 

\begin{itemize} 
\item{} Total mass about $10^{15}$ solar masses, mostly dark matter of 
unknown nature. Dark matter seems to be non-interacting, except through 
gravitation. 

\item{} Gas mass about $10^{14}$ solar masses. This gas is enriched, 
and so has been reprocessed through stars to some degree, since it is 
often about 1/3 of solar abundances. The temperature of the gas is 
between 5 - 10 keV, or up to about $10^8$ K. The density of the gas at 
the center is about $3 \times 10^{-4}$ cm$^{-3}$. The X-ray luminosity 
from thermal bremsstrahlung emission of the hot gas is about $10^{44}$ 
erg/s, and the energy content in the gas is about $10^{62}$ erg. 

\item{} The total stellar mass is considerably less than the gas 
mass. 

\item{} The radial scale of the density profile of the cluster is about 
300 kpc. The ``outer edge" is at a few Mpc. This is where the accretion 
shock becomes visible through reacceleration of an old quiescent 
population of energetic particles, and subsequent radio emission 
(En{\ss}lin et al. 1998). 

\item{} The cooling time of the gas is usually larger than the Hubble 
time, of about $5 \times 10^{17}$ s. There are some clusters for which 
the central cooling time is shorter than the Hubble time and for those a 
cooling flow is inferred. This cooling flow can reach some $10^3$ solar 
masses per year in accretion, of which few traces are visible anywhere. 
XMM-Newton data suggest that much of this cooling is compensated by 
heating, to be discussed below, and as had been suspected for some time; 
references are given below. 

\end{itemize} 

Groups of galaxies 
(e.g., Biermann et al. 1982, Biermann \& Kronberg 1983, 
Biermann \& Kronberg 1984a, Biermann \& Kronberg 1984b, 
Schmutzler \& Biermann 1985, Wu \& Xue 2001), and also 
the large sheets and filaments in the large scale distribution of 
galaxies also appear to harbor much gas, often at lower temperature than 
in rich clusters, and so is very difficult to detect. This gas is 
suspected to constitute the missing baryonic matter, inferred from 
comparing nucleosynthesis results and microwave background fluctuation 
data with the baryonic mass directly detectable in stars. 

\subsection{Magnetic fields} 

Here we discuss the recent detection of rather strong magnetic 
fields in clusters (Clarke et al. 1999, Clarke et al. 2000). 

Magnetic fields have now been firmly detected in clusters of galaxies, and 
as a rule appear to have the following properties: 

\begin{itemize} 

\item{} Strength at or above 5 - 10 $\mu$G normally. 

\item{} In two known cases the magnetic field strength reaches 
equipartition, with 20 - 30 $\mu$G (see, e.g., En{\ss}lin et al. 1997). 

\item{} The reversal scale is small, 10 kpc or less. The field is 
highly chaotic. 

\item{} The scale of the magnetic region in the cluster gas has a radius 
of at least about 500 kpc. This means that the magnetic field is 
detectable through Rotation Measure data throughout the inner one Mpc. 

\end{itemize}

\section{Sources of cosmic rays in clusters} 

Here we discuss the various possible sources of cosmic rays in clusters 
of galaxies.

\subsection{Cosmic rays: Starburst and normal galaxies} 

Here we discuss cosmic rays in clusters, as derived from normal galaxies 
and starburst galaxies (Popescu et al. 2000, V{\" o}lk et al. 1999, V{\" 
o}lk et al. 2000). 

Galaxies and especially starburst galaxies produce an 
enormous power in cosmic rays, as evidenced through the far-infrared/radio 
correlation. The far-infrared luminosity is therefore a direct measure of 
the creation of cosmic rays, 

\begin{equation} 
L_{CR} \; \simeq 10^{-3} \, L_{FIR}. 
\end{equation} 

These cosmic rays are generally believed to drive a wind from the galaxy, 
and so lose their power in adiabatic expansion to the wind. These cosmic 
rays can then be captured by the environment in a cluster. These cosmic 
rays then outside have little total power left, only a small fraction of 
their original power, but lots of ionizing capital (Nath et al. 1993). 
However, Volk, Aharonian, \& Breitschwerdt (1996) argued that the 
termination shock waves of such galactic winds can re-accelerate the 
escaping cosmic rays. 

Since galaxies in clusters are usually early Hubble type galaxies 
their far-infrared luminosity is relatively weak, and so their starburst 
activity and a fortiori cosmic ray power is diminished 
(Popescu et al 2000, V{\" o}lk et al. 1999, V{\" o}lk et al. 2000). 

If the galaxies get stripped on their path through the cluster (e.g., 
Himmes \& Biermann 1980, Toniazzo \& Schindler 2001), 
then their cosmic ray contribution may not suffer 
from driving any wind, and so their contribution may be larger. 

However, on balance the cosmic ray power from galaxies is not expected to 
dominate over other sources of cosmic rays in clusters of galaxies.

\subsection{Cosmic rays: Accretion shocks} 

Here we discuss the cosmic rays from accretion shocks (Kang et al. 1997, 
En{\ss}lin 1998). 

Accretion shocks around clusters of galaxies, now detected 
through the re-energization of fossil clouds of energetic particles 
(En{\ss}lin et al. 1998), can certainly also accelerate cosmic rays, and 
do so to very high particle energy (Kang et al. 1997). These 
accretion shocks, however, sample the potential well at a large distance, 
several Mpc, and so cannot contribute very large power. 

There is probably a similar feature around filaments and sheets in the 
galaxy distribution (En{\ss}lin et al. 2001), 
in one example detected through the 
renergization of an extended radio jet. 

These arguments strengthen the case for the validity of the cosmic 
simulations, which show shocks around clusters, filaments and sheets in 
the galaxy distribution. 

%%%%%%%%%%%%%%%%%% figure%%%%%%%%%%%%%%%%%%% 
\begin{figure} 
\centering 
\vspace{0cm} 
\epsfig{file=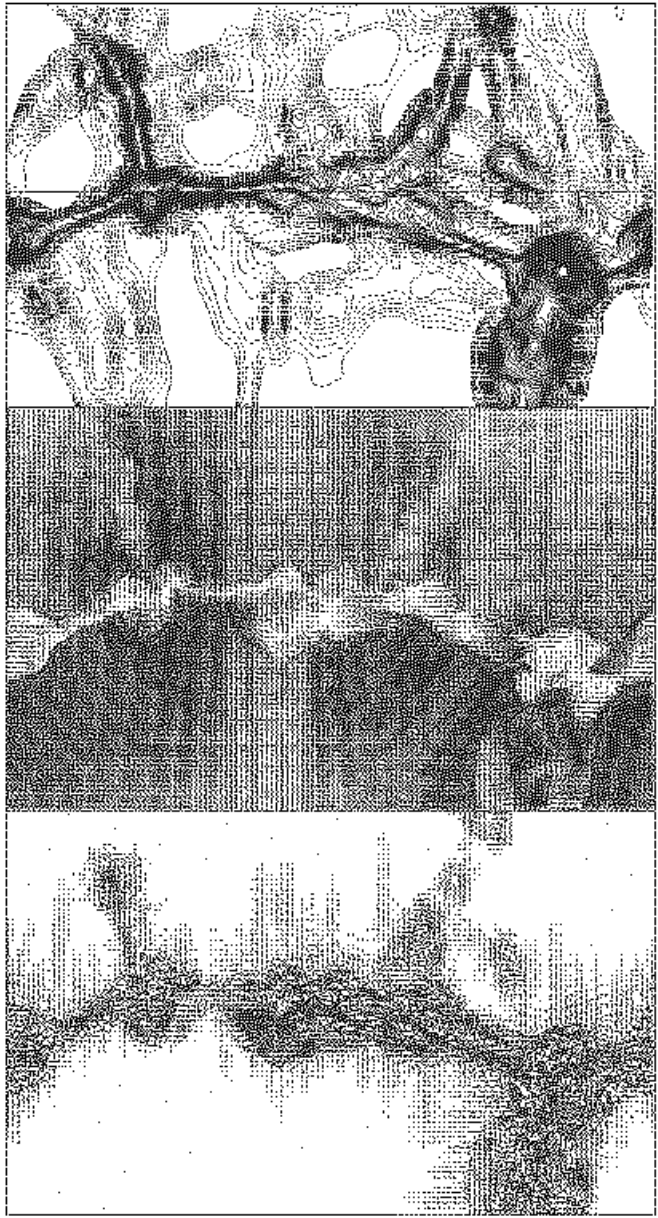,width=13cm} 
\vspace{0cm} 
\caption{Two-dimensional cut of the simulated universe at $z=0$ 
from a cosmological simulation. 
The plot shows a region of $32h^{-1}\times20h^{-1}{\rm Mpc}^2$ with 
a thickness of $0.25 h^{-1}{\rm Mpc}$, although the simulation was 
done in a box of $(32 h^{-1}{\rm Mpc})^3$ volume. 
The first panel shows baryonic density contours, the second panel 
shows velocity vectors, and the third panel shows magnetic field vectors. 
In the third panel, the vector length is proportional to the log of 
magnetic field strength. 
(Ryu et al. 1998)} 
\end{figure} 
%%%%%%%%%%%%%%%%%%%%%%%%%%%%%%%%%%%%%%%%%%%% 

%%%%%%%%%%%%%%%%%% figure%%%%%%%%%%%%%%%%%%% 
\begin{figure} 
\vspace{0cm} 
\epsfig{file=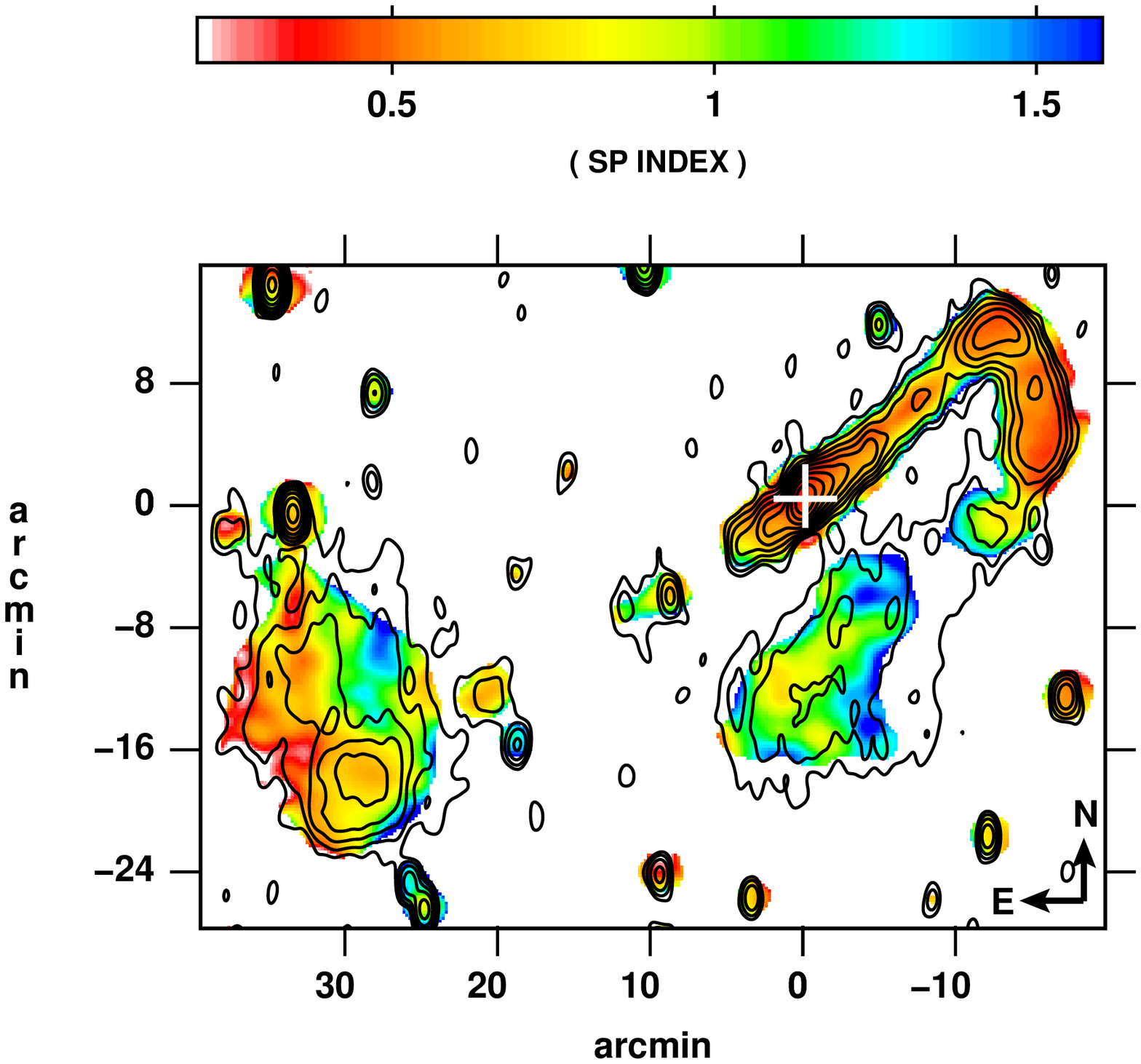,width=12cm} 
\epsfig{file=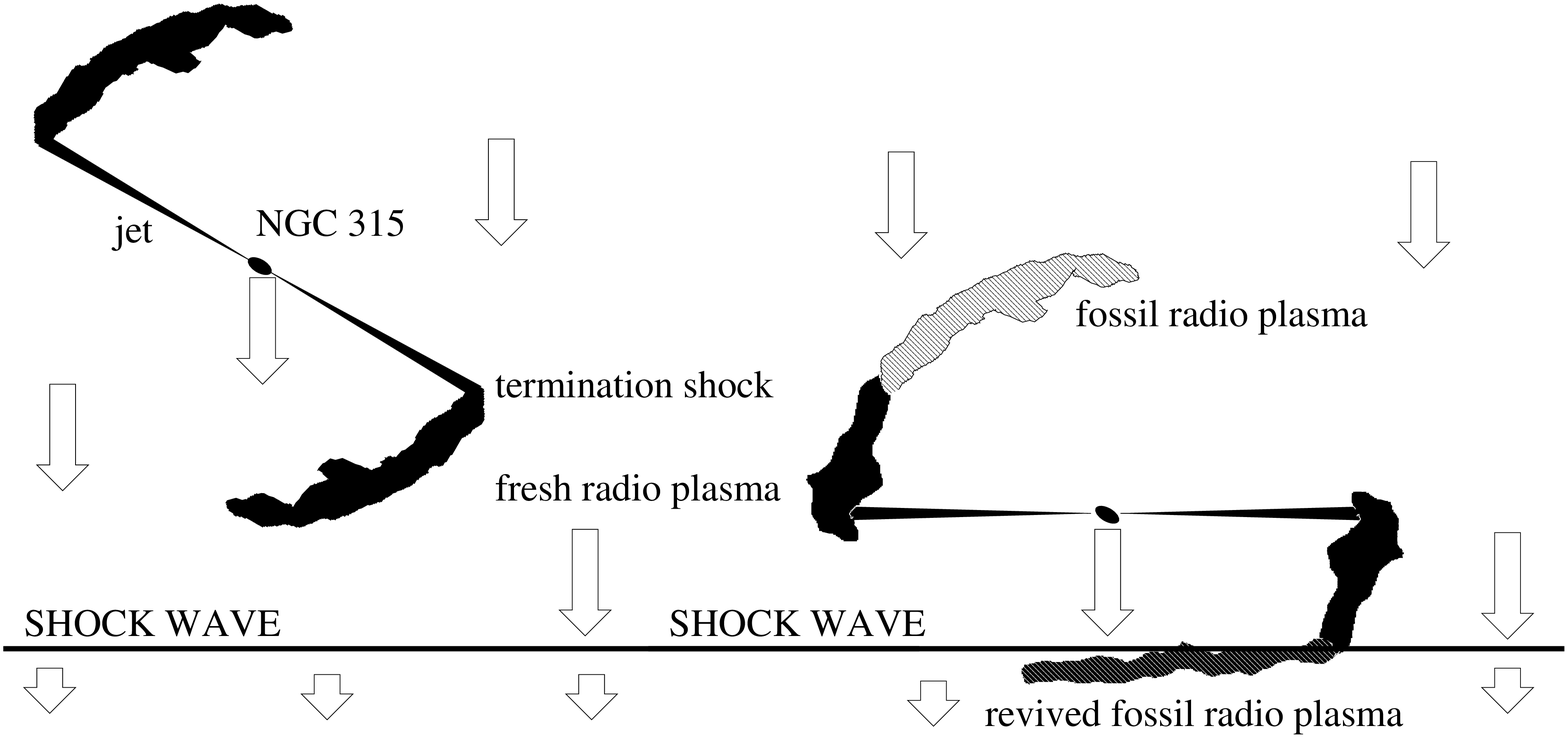,width=12cm} 
\vspace{0cm} 
\caption{Here we show how we might interpret the configuration 
around the radio galaxy NGC315, as influenced by the accretion shocks 
around the local filaments. The top figure shows the peculiar radio 
morphology of the radio galaxy. The western radio trail exhibits a 
sharp bending and a flat (color-coded) spectral index. This can be 
understood as signatures of a environmental shock wave, as sketched in 
the bottom figure (En{\ss}lin et al. 2001).} 
\end{figure} 
%%%%%%%%%%%%%%%%%%%%%%%%%%%%%%%%%%%%%%%%%%%% 

\subsection{Cosmic rays: Mergers and relaxation} 

Mergers between clusters and structural rearrangements of un-relaxed 
clusters can also produce shock waves and so cosmic rays (Donnelly et al. 
2001, Schindler 2002). 

When two clusters merge, as e.g. in the case of the Perseus cluster, they 
rearrange themselves over some time - about one to two crossing times, 
causing widespread shock waves and highly anisotropic phase and real space 
distributions of galaxies (e.g. Donnelly et al. 2001, Schindler 2002). 
The group around NGC383 is another example where the group appears as a 
cigar in projection on the sky, contains only early Hubble type galaxies, 
and the radio galaxy 3C31 (the host galaxy of which is NGC383 itself) 
shoots its radio jets along the cigar figure in projection. Again cosmic 
rays can be accelerated by the shocks caused by mergers, but as these 
shocks are low Mach number, the spectrum of the accelerated particles is 
steep, i.e. will not contribute much at high energy. On the other hand, 
their total energy content could approach the thermal energy.

\subsection{Cosmic rays: Radio galaxies} 

Finally, cosmic rays from radio galaxies can surely provide sufficient 
energy to provide cosmic rays, and also magnetic fields to clusters of 
galaxies, and do so to the stability limit of the medium (En{\ss}lin, 
et al. 1997). 

Radio galaxies can easily provide lots of energy in relativistic 
particles and also in magnetic fields, all derived from the accretion 
power of their central black hole. The jet-disk symbiosis picture 
developed by Falcke over the past ten years 
(Falcke \& Biermann 1995, Falcke et al. 1995a, Falcke et al. 1995b, 
Falcke \& Biermann 1999) can be used to derive the 
total power injected by radio galaxies over time, and can easily be seen 
to surpass all the thermal energy in the cluster, which obviously taps 
the gravitational potential well. This means that radio galaxies can 
push the thermal stability of the cluster gas to its limit, thus at least 
in principle ensuring to keep it at threshold of stability, just as has 
been argued for the interstellar medium. In the case of Hydra A this 
limit appears to be reached, because the nonthermal bubble has broken 
through to the outside (McNamara et al, 2000, David et al. 2001, 
Nulsen et al. 2001). What is not so clear 
is whether radio galaxies can do this frequently enough in time to 
continuously keep the intergalactic medium at the threshold of stability 
with respect to the content in magnetic fields and in energetic particles. 

%%%%%%%%%%%%%%%%%% figure%%%%%%%%%%%%%%%%%%% 
\begin{figure} 
\vspace{0cm} 
\epsfig{file=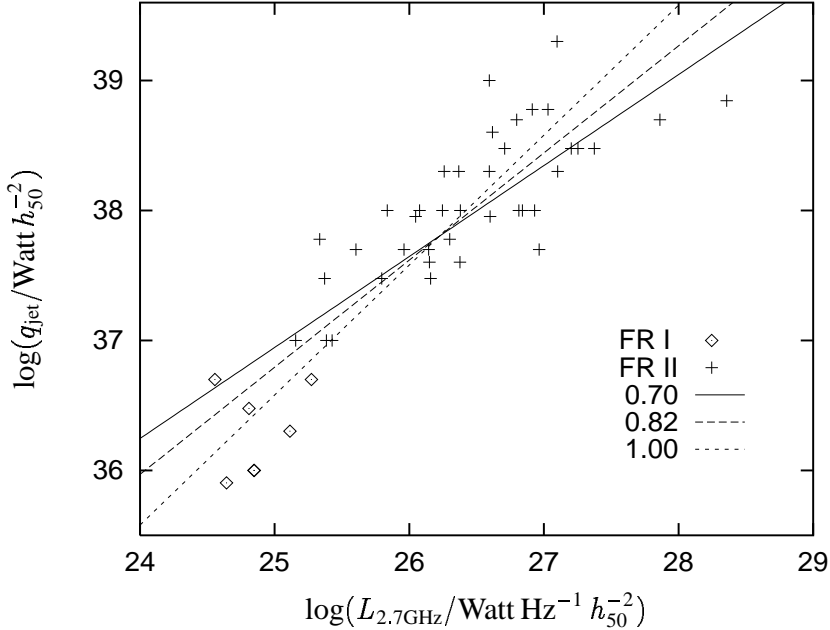,width=11cm} 
\vspace{0cm} 
\caption{Jet-power versus the radio emission 
for radio galaxies, demonstrating the enormous power residing in the jet 
output. (En{\ss}lin et al. 1997)} 
\end{figure} 
%%%%%%%%%%%%%%%%%%%%%%%%%%%%%%%%%%%%%%%%%%%% 

%%%%%%%%%%%%%%%%%% figure%%%%%%%%%%%%%%%%%%% 
\begin{figure} 
\vspace{0cm} 
\epsfig{file=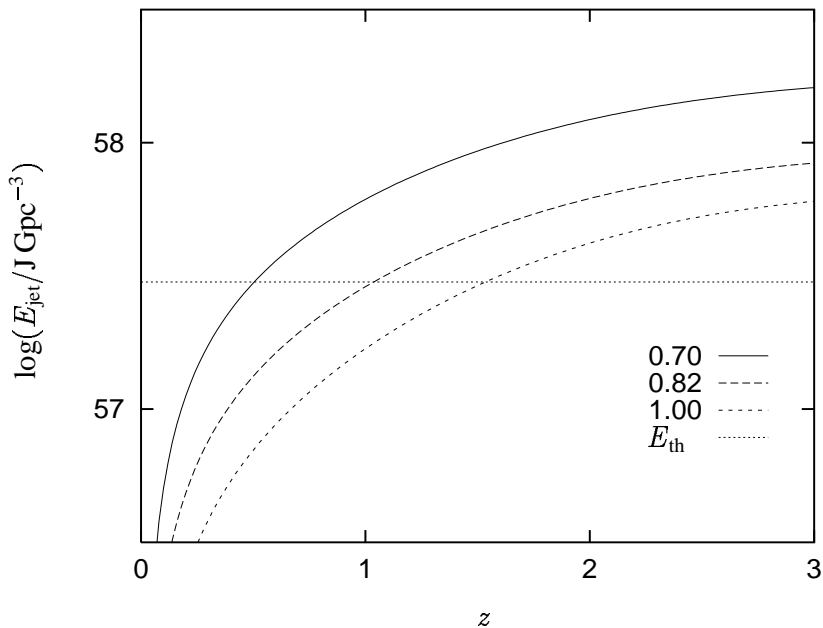,width=11cm} 
\vspace{0cm} 
\caption{Integrated jet-power versus time 
in a cluster of galaxies, showing the total energy as compared to the 
thermal energy. (En{\ss}lin et al. 1997)} 
\end{figure} 
%%%%%%%%%%%%%%%%%%%%%%%%%%%%%%%%%%%%%%%%%%%% 

\section{Magnetic fields in the cosmos} 

Then, we describe the attempts to understand where cosmic magnetic 
fields come from. 

Magnetic fields are known to exist almost everywhere in the cosmos, with 
the Earth and the Sun the most prominent examples 
(Kronberg 1994, Beck et al. 1996, Kulsrud et al. 1997, Blasi et al. 1998, 
Kulsrud 1999, Blasi \& Olinto 1999). The Sun 
demonstrates that a dynamo mechanism can twist, fold, and turn back on 
itself magnetic loops and so strengthen the magnetic field, a mechanism 
proposed by Steenbeck \& Krause 
(Steenbeck \& Krause 1965, Steenbeck et al. 1966, Steenbeck \& Krause 
1966, Krause \& Steenbeck 1967, Steenbeck et al. 1967, 
Krause 1967, Steenbeck \& Krause 1969a, 
Steenbeck \& Krause 1969b, R{\"a}dler 1968a, R{\"a}dler 1968b, 
R{\"a}dler 1969a, R{\"a}dler 1969b, R{\"a}dler 1970, Krause 1969) as 
well as Parker (Parker 1969, Parker 1970a, Parker 1970b, Parker 1970c, 
Parker 1971a, Parker 1971b, Parker 1971c, Parker 1971d, Parker 1971e, 
Parker 1971f, Lerche \& Parker 1971, Lerche \& Parker 1972, 
Parker 1973, Parker 1975a, Parker 1975b) 
many years ago. This mechanism requires a seed field, and a simple 
plasma physics mechanism was proposed by L. Biermann 
(Biermann 1950, Biermann \& Schl{\"u}ter 1951) more than fifty years ago; 
this is based on the idea 
that in the equation of motion for electric currents the non-coincidence 
between the surfaces of constant pressure and the surfaces of constant 
density drives a current, which cannot be compensated by any electric 
field; this is normally expected for almost any rotating system such as a 
galaxy or a star. This provides a seed field, which is very weak, and 
any normal dynamo mechanism requires many rotation periods to reach any 
interesting strength for the magnetic field. In the Sun this is believed 
to be possible at the base of the convective layer (Cowling 1953, Mestel 
\& Roxburgh 1962), as in massive star cores; massive stars 
have been observed to show in some examples nonthermal radio emission, 
clearly demonstrating the existence of non-negligible magnetic fields on 
the radiative surface, and at the base of their winds 
(Biermann \& Cassinelli 1993, Seemann \& Biermann 1997). 
In a galaxy there may not be sufficient time to develop appreciable 
magnetic fields by this mechanism. 

The observations of galaxies, however, indicate strong constraints for 
any mechanism to strengthen magnetic fields: 

\begin{itemize} 

\item{} In our Galaxy the time scale of ``turnover" of the interstellar 
medium as derived from cosmic ray studies is about 30 million years; 
this is the time scale to replenish the cosmic ray population. This means 
that any geometric order in the homogeneity of the magnetic field is 
destroyed with such a time scale, and yet, the magnetic field is of order 
half ordered. 

\item{} The observation of starburst galaxies that also have 
characteristic time scales of also a few tens of millions of years, and 
yet with an ordered magnetic field in near equipartition with the 
interstellar medium, again indicates a very fast mechanism to strengthen 
and so regenerate the magnetic field. 

\end{itemize} 

This means that the mechanism, whatever it may be, works basically on the 
Alfv{\'e}nic time scale through the thickness of the hot disk 
(Kaneda et al. 1997, Snowden et al. 1997). 
This is also the limiting time scale for any 
inverse cascade model, which may be hard to reach 
(Stribling \& Matthaeus 1991, Stribling \& Matthaeus 1994, 
Stribling \& Matthaeus 1995, Matthaeus et al. 1996, 
Matthaeus et al. 1998, Matthaeus et al. 1999). 

Once the magnetic fields in galaxies are understood 
(Han \& Qiao 1994, Han et al. 1997, Han et al. 1997, 
Krause \& Beck 1998), the universe can be ``filled" with 
magnetic fields and cosmic rays 
(Ryu et al. 1998, En{\ss}lin et al. 1998, Kronberg et al. 1999, 
Birk et al. 2000, Kronberg et al. 2001).

\section{Explosions in a homogeneous medium} 

We then discuss explosions into a homogeneous medium, in the spirit of 
Kellermann \& Pauliny-Toth (1968), and McKee \& Ostriker (1977) and so as 
to describe the effective cooling in either the interstellar medium or 
intergalactic medium, subject to the buffeting of supernova explosions or 
radio galaxy explosions. 

We also discuss to what limit this analogy may help to understand the 
physics of these two media. Also, the recent XMM-Newton findings 
illustrate the necessity to consider the heating from active galaxies 
in 
clusters. 

In this section we first wish to consider the inhomogeneity of the 
intergalactic medium, just as implied already by the irregular explosions 
that take place, and so fully explore the analogy with the interstellar 
medium. 

Following the treatment of explosions in the interstellar medium 
originally proposed by McKee \& Ostriker (1977) we wish to derive here 
a crude estimate of the effective cooling rate of medium riddled with 
explosions. The line of reasoning also follows the argument for the 
Compton catastrophe in Active Galactic Nuclei 
(Kellermann \& Pauliny-Toth 1968). 

Consider first a homogeneous medium, subject to aperiodic explosions which 
carve out a spherical bubble, surrounded by a shell of density four times 
higher, caused by the strong shock. We adopt the adiabatic limit, and so 
one can describe these explosions approximately by a self-similar Sedov 
solution. Then the shock velocity $v_{sh}$ scales with radius $r$ as 
$r^{-3/2}$, while the radius itself decreases with time $t$ as 
$t^{-2/5}$. The shell has a thickness of $\Delta r = r/12$, and has a 
density relative to the reference density $n_0$ of $4 n_0$. We then 
consider multiple explosions, which overlap, and so in a very simple 
description have from a first level $1/4$ the volume with $4$ times the 
density, and so this contributes to the thermal X-ray emission in 
incremental X-ray luminosity 

\begin{equation} 
\Delta L_{X,1} \sim \frac{1}{4} \; (4)^2 \; f(T_1). 
\end{equation} 

From the second and higher level $n$ the contribution is 

\begin{equation} 
\Delta L_{X,n} \sim (\frac{1}{4})^n \; (4)^{2 n} \; f(T_n). 
\end{equation} 

At each level the emission coefficient corresponds to the local 
temperature $T_n$, but as the emissivity varies rather weakly with 
temperature compared to the dependence of emission on density, we can 
approximate this as a general average emission coefficient $f(T_{av})$, 
and write 

\begin{equation} 
\Delta L_{X,\Sigma} \sim Vol \, n_0^2 \, \left( 1 + \frac{1}{4} \; 
(4)^2 \, f(T_1) \, + \, (\frac{1}{4})^2 \; (4)^{4} \, f(T_2) \, + {\rm 
...} \right), 
\end{equation} 

which is 

\begin{equation} 
\Delta L_{X,\Sigma} \sim Vol \, n_0^2 \, f(T_{av})?, \Sigma 
(\frac{1}{4})^m \; (4)^{2 m}. 
\end{equation} 

The sum is broken off as soon as the local cooling time becomes shorter 
than the repetition time scale for explosions. This may be the case 
already after the second level. The second level increases the 
emission by more than an order of magnitude. 

The energy content in the same approximation is modified as follows 

\begin{equation} 
Vol \, n_0?, k_B \, \Sigma (\frac{1}{4})^m \; (4)^{m} T_m \; \approx 
Vol \, n_0?, k_B \, T_{av}, 
\end{equation} 

where $k_B$ is the Boltzmann gas constant. 
Therefore the total X-ray luminosity may be increased by the strong 
inhomogeneity of the medium easily by a factor of order ten. Applying 
this to the interstellar medium the cooling time inferred from the 
assumption of homogeneity using the numbers of Snowden et al. (1997), 
Kaneda et al. (1997), Valinia \& Marshall (1998), of a density of $3 
\times 10^{-3}$ per cm$^3$, and a temperature of $4 \times 10^6$ K, 
we find a cooling time of about $3 \times 10^8$ years, and then using the 
argument on inhomogeneity derived here, the cooling time may well be very 
much shorter, and could be only $3 \times 10^7$ years. Interestingly, 
this suggests that the time scales for heating and cooling of the 
hot medium of the interstellar gas, the time scale for magnetic 
field regeneration, and also the time scale for cosmic ray 
transport may all be the same. This then would lead to an 
understanding why we may on average always be at a limit of 
stability. Surely these considerations may apply to the present day 
interstellar medium. On the other hand, we have to realize, that these 
arguments may be less important in today's clusters of galaxies due to the 
scarcity of active phases of galaxies, but they were probably critical in 
the early evolution, when the activity level in the Universe was very much 
higher, by about a factor of 30 near redshift near 1.7 (e.g., Pugliese 
\etal 2000); some clusters of galaxies are known to harbour two radio 
galaxies, even today.

\subsection{Reheating of gas in clusters of galaxies} 

Repeating the same argument for clusters of galaxies the 
cooling time may be shorter than the Hubble time in many cases, 
especially locally, but would then just lead to a refueling of an Active 
Galactic Nucleus, which then through the action of its jet, reheats the 
local environment; and we know from the work of Falcke 
\etal (Falcke \& Biermann 1995, Falcke et al. 1995a, 
Falcke et al. 1995b, Falcke \& Biermann 1999) that the jet carries a 
power equivalent to the visible electromagnetic radiation, and so easily 
can provide substantial heating. Also, the recent XMM-Newton findings 
illustrate the necessity to consider the heating from radio galaxies in 
the cores of clusters (Churazov et al. 2002, B{\" o}hringer et al. 2002, 
Br{\" u}ggen et al. 2002). As almost all galaxies are now known to 
harbor a central black hole, this threshold argument should work with 
most galaxies in a cluster.

\section{Cosmological structure of magnetic fields} 

Here we describe some cosmological simulations that show what happens 
with magnetic fields (Biermann et al. 1997, Ryu et al. 1998). We then 
explore the topology of magnetic fields in a medium subject to continuous 
excitement of turbulence, again as an example for both the interstellar and 
the intergalactic medium (Lee et al. 2002). 

As shown in the large scale simulations by Ryu et al. 
(Biermann et al. 1997, Ryu et al. 1998), the baryonic flow 
in the large scale structure 
formation draws the magnetic fields along, so as to reproduce the galaxy 
distribution also in the magnetic field distribution. The magnetic field 
is strongest in clusters of galaxies, somewhat weaker in the filaments, 
even weaker in the sheets, and very weak in the voids. The flow is along 
the filaments and sheets towards the next cluster or supercluster, 
pulling the magnetic fields along. The clusters, filaments and sheets 
are all bounded by shocks; the existence of these shocks is supported by 
observational data (En{\ss}lin et al. 1998, En{\ss}lin et al. 2001).

\subsection{Topology of magnetic fields} 

In any otherwise homogeneous medium continuously excited by turbulence 
the magnetic field can form long flux tubes, depending on the 
intermittency, a measure of the Alfv{\'e}nic Mach number of the 
turbulence injection. The weaker the magnetic field on average, the 
stronger the inhomogeneity of its distribution. 

%%%%%%%%%%%%%%%%%% figure%%%%%%%%%%%%%%%%%%% 
\begin{figure} 
\vspace{0cm} 
\epsfig{file=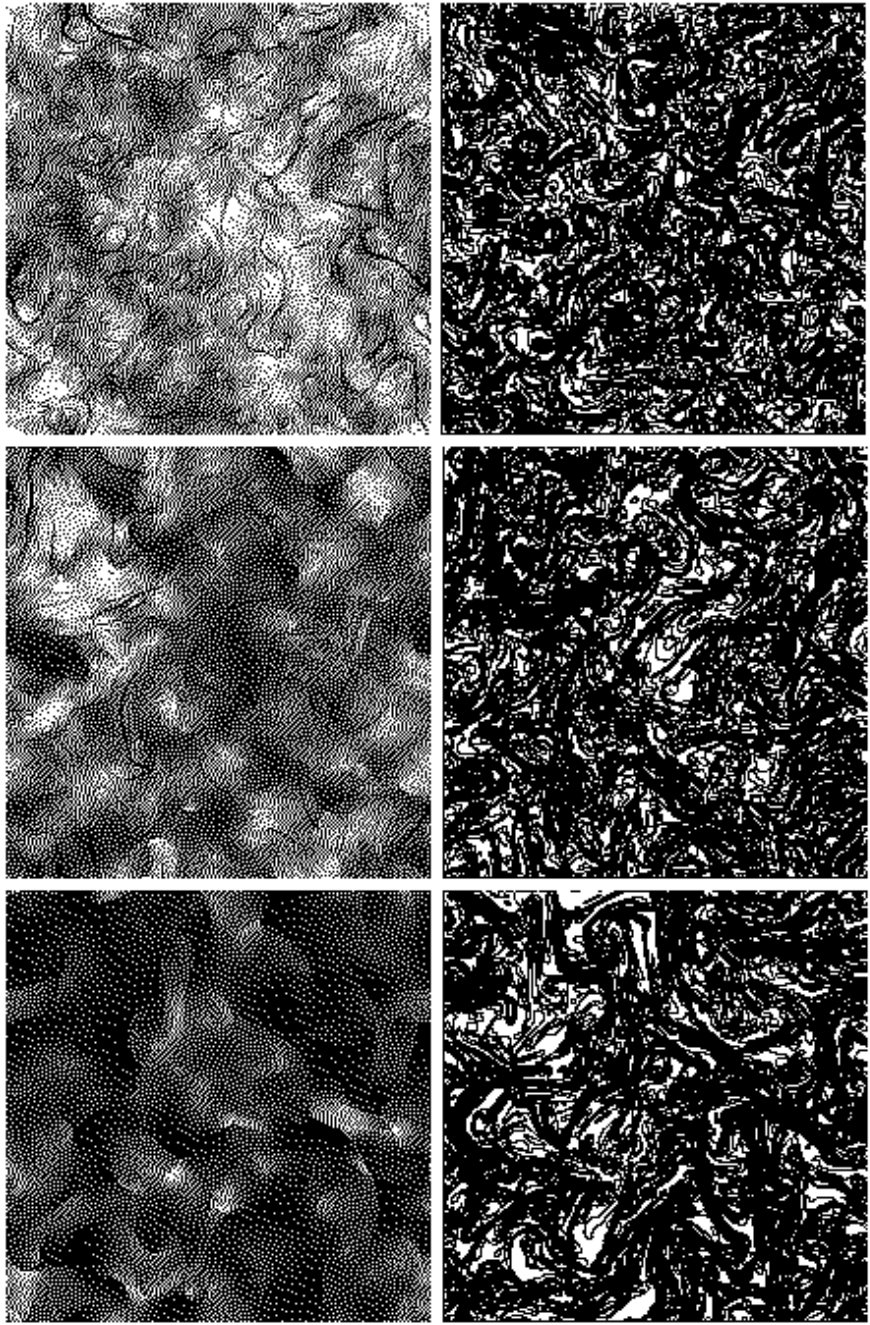,width=16cm} 
\vspace{-1cm} 
\caption{Grey scale images of density (left panels) and contours 
of magnetic field lines (right panels) at an epoch 
from simulations with various field strength of intermittent 
turbulence in a 2D case, whith a magnetic field, stretched along into 
long flux tubes. 
In the density images, brighter regions represent higher values and 
the gray scale has been set arbitrarily to highlight structures. 
(Lee et al. 2002)} 
\end{figure} 
%%%%%%%%%%%%%%%%%%%%%%%%%%%%%%%%%%%%%%%%%%%% 

Therefore we have to consider a highly inhomogneoeus medium in density, 
in magnetic fields, and probably also in cosmic rays. And yet, on 
average the time scales for their supply and decay are so close to each 
other, that an apparent regularity is reached.

\section{Lepton energy distribution} 

We furthermore discuss the two sources of leptons so as to predict the 
lepton energy distribution, critical to understand the inverse-Compton and 
synchrotron emission in clusters; leptons are expected both from the 
primary acceleration of electrons starting from the thermal tail, as well 
as from pion decay following p-p collisions between energetic protons and 
the medium itself (Falcke \& Biermann 1995, Falcke et al. 1995a, Falcke et 
al. 1995b, Falcke \& Biermann 1999, En{\ss}lin \& Biermann 1998). 

There are two basic modes of lepton energy distribution; we consider two 
modes using diffusive shock acceleration as an example to illustrate: 

First, those electrons accelerated outwards in phase space from 
the thermal tail, i.e. starting with a power law from somewhere near the 
peak of the Maxwellian; this Maxwellian itself could be relativistic, 
since electrons get thermally relativistic from temperatures near 
$10^{10}$ K already. Also, this power law might have several different 
slopes, as a function of energy, in response to losses, or in response to 
changes of the diffusion coefficient as a function of energy in the 
acceleration region. This latter effect might happen if the electrons at 
first sample an electric field in the case that their Larmor radius is 
actually smaller than the thermal proton Larmor radius, which is likely 
to define the thickness of a shock (and then with the post-shock 
temperature). The first effect might happen if synchrotron and inverse 
Compton losses cut on immediately after the acceleration is finished, or 
even inside the acceleration region. Also, if the shock is relativistic 
itself, further modifications to the spectrum might exist (e.g. Bednarz 
\& Ostrowski 1998). 

Second, the electrons might be the decay products from pion decay, and in 
this case there would be an equal number of positrons; if neutrons 
are also produced in collisions, and their subsequent decay is far away in 
distance, then locally the electrons and positrons would differ slightly 
in number density. In this second case the energy distribution starts 
near the pion mass, and is strongly suppressed at lower energies (only 
inelastic collisions fill this part of phase space, an effect which is 
well known from anti-protons). Following pion decay the peak of the 
energy in the population is near pion mass in most cases, as opposed to 
the thermal case (above), where that peak is normally near the rest mass 
of the electron (i.e. for spectral indices near to or steeper than -2 in 
the energy distribution, written in transrelativistic momentum $pc$: $4 
\pi \, (pc)^2 \, f(pc)$, where f(pc) is the 3D distribution function). As 
a consequence, for instance, given the same energy content in the lepton 
population, the radio emission may be very different between the two 
cases, since the electrons responsible for radio emission usually are 
those far above either threshold, but have a very different flux. 

Obviously, adiabatic losses can shift these spectra around in phase 
space, synchrotron and inverse Compton losses can truncate them at the top 
energies, while ionization losses can truncate or weaken it at the lower 
energies, all subsequent to acceleration. 

In earlier work we have argued that radio galaxies typically use the 
second mode, and derive all the leptons visible through radio synchrotron 
emission from pion decay. The pions are created near the base of the jet 
(Biermann et al, 1995, Falcke \& Biermann 1995, Falcke et al. 1995a, 
Falcke et al. 1995b, Falcke \& Biermann 1999). 

%%%%%%%%%%%%%%%%%% figure%%%%%%%%%%%%%%%%%%% 
\begin{figure} 
\vspace{0cm} 
\epsfig{file=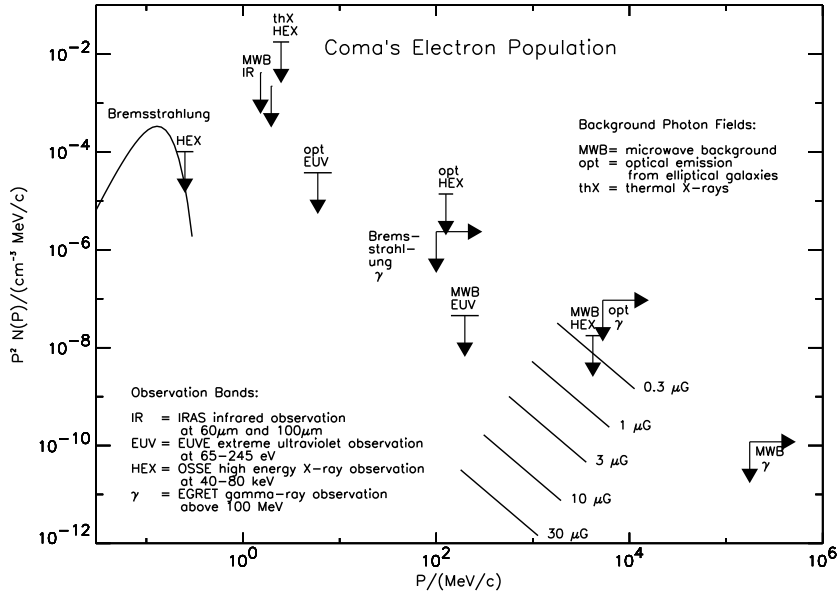,width=12cm} 
\vspace{0cm} 
\caption{Upper limits to the electron and 
positron energy distribution for the Coma cluster. 
(En{\ss}lin \& Biermann 1998)} 
\end{figure} 
%%%%%%%%%%%%%%%%%%%%%%%%%%%%%%%%%%%%%%%%%%%% 

\section{Tests} 

High spatial resolution Rotation Measure data should give a clue on 
intermittency and Alfv{\'e}n number of turbulence, a key ingredient in 
understanding both the interstellar medium and the intergalactic medium. 

Comparison of soft/hard X-rays, line emission and absorption should show 
whether we are in the convective or diffusive limit for cosmic ray 
transport and, consequently, will delimit the effect of cosmic ray 
heating. $\gamma$-rays will provide another test, as will the connection 
to low frequency radio emission (e.g. En{\ss}lin et al. 1997, En{\ss}lin 
\& Biermann 1998, Br{\" u}ggen et al. 2002).

\section{Summary} 

Clusters of galaxies probably have a chaotic relaxation cycle with their 
input of thermal energy, magnetic energy, and cosmic ray energy; we 
predict that radio galaxies provide the bulk, when considered the long 
term effect. 

The maximum of the sum of the energy density of magnetic fields and cosmic 
rays is probably near equipartition with the thermal gas, due to the 
relaxation cycle probably being extended to a factor of order ten below 
equipartition. 

Energy input may occur from radio galaxies at random rare times, mergers, 
the accretion shocks, and possibly even starbursts in merged galaxies. 

So there are two limits that we consider to be possible. 
Consider the energy distribution: 

\begin{equation} 
E_B + E_{CR} \; \simeq \; (\frac{1}{10} {\rm ....} 1) \, E_{th} 
\end{equation} 

where $E_B$, $E_{CR}$, and $E_{th}$ are the total energy content in the 
cluster intergalactic medium for magnetic fields, cosmic rays, and the 
thermal energy. 

\noindent 1) The evolution of the ratio of the two sides is highly 
time dependent, with occasional deep dips, or 

\noindent 2) The ratio is only slowly varying. 

\noindent Obviously, considering the dependencies on location in the gas 
this ratio will be extremely time-dependent, and vary by powers of ten. 

The extreme cases of complete equipartition, and the case of highly 
intermittent injection of cosmic rays and magnetic fields, along with the 
heating provided by jets, in a mode where the intergalactic medium is 
pushed to its stability limit, might be well worth exploring. Also, the 
case, when the magnetic field is relatively weak at injection, but then 
stretched into long flux tubes that enhance its effect, should be 
considered. The medium is also likely to be extremely inhomogeneous, 
with filaments and folded sheets of higher density, different 
temperature, and different cosmic ray particle density. And yet, in all 
its complexity, it is plausible that simple rules govern its {\it 
average} behavior. Such is suggested by the radio-far infrared 
correlation of the thermal dust emission with the non-thermal radio 
emission, and the presumably thermal X-ray emission from normal galaxies 
and starburst galaxies. Taking radio galaxy explosions in place of 
supernova explosions, and considering clusters of galaxies in place of the 
interstellar medium, it appears plausible that similar simple 
relationships govern the intergalactic gas - we just need to discover 
them.

\acknowledgements 

PLB appreciates the diligent work of making suggested improvements to the 
manuscript by the editor St. Bowyer, and my colleague P.P. Kronberg. 
PLB would like to thank P.P. Kronberg and T. W. Jones for many years of 
very inspiring discussions. High energy work with PLB is funded through 
grant 05 CU1ERA/3 from DESY/BMBF. The collaboration between HK, DR and 
PLB has been funded by the DFG in Germany and KOSEF in Korea.

\end{document}